\documentclass[amsmath,amssymb,pre,onecolumn,tightenlines,superscriptaddress,floatfix,final]{revtex4-2}
\usepackage{bm}
\usepackage{nicefrac}
\usepackage{graphicx}
\usepackage{dcolumn}
\usepackage{bm}
\usepackage[breaklinks,colorlinks = true,linkcolor = blue,urlcolor=blue,citecolor=blue]{hyperref}
\usepackage[dvipsnames]{xcolor}
\usepackage{soul}
\makeatletter
\newcommand{\rc}[1]{\@ifclasswith{revtex4-2}{final}{#1}{\textcolor{olive}{#1}}}
\makeatother

\begin{document}

\title{The dynamics of thermalisation in the Galerkin-truncated, three-dimensional Euler equation}

\author{Rajarshi}
\email{rajarshic.edu@gmail.com}
\affiliation{International Centre for Theoretical Sciences, Tata Institute of Fundamental Research, Bangalore 560089, India.}
\author{Mohammad Saif Khan}
\email{mohammad.saifkhan@icts.res.in}
\affiliation{International Centre for Theoretical Sciences, Tata Institute of Fundamental Research, Bangalore 560089, India.}
\author{Prateek Anand}%
\email{prateekanand85@gmail.com}
\affiliation{Department of Mechanical Engineering, Indian Institute of Technology Bombay, Powai, Mumbai 400076, India.}
\author{Samriddhi Sankar Ray}
\email{samriddhisankarray@gmail.com}
\affiliation{International Centre for Theoretical Sciences, Tata Institute of Fundamental Research, Bangalore 560089, India.}

\begin{abstract}
    The inviscid, partial differential equations of hydrodynamics when projected
	via a Galerkin-truncation on a finite-dimensional subspace spanning
	wavenumbers $-{\bf K}_{\rm G} \le {\bf k} \le {\bf K}_{\rm G}$, and
	hence retaining a finite number of modes $N_{\rm G}$, lead to absolute equilibrium states. We
	review how the Galerkin-truncated,
	three-dimensional, incompressible Euler equation thermalises and its  
	connection to questions in turbulence. We
	also discuss an emergent pseudo-dissipation range in the energy spectrum and  
	the time-scales associated with thermalisation.
\end{abstract}

\maketitle
\section{Introduction}

Turbulent flows, with its tell-tale self-similar, intermittent, chaotic and
	multifractal signatures, \rc{ 
	describe phenomena spanning from large scale winds, down to small scale dynamics of swimming bacteria \cite{Vallis06,mukherjee2023}}.  For
	classical turbulence, such flows are solutions of the Navier-Stokes
	equation, with appropriate boundary conditions, forcing and small
	enough viscosities $\nu$, leading to driven-dissipative,
	non-equilibrium stationary states which defy a Hamiltonian description.
	This lack of a Hamiltonian description, among others, makes the problem
	of turbulence a particularly difficult one, especially when viewed through
	the lens of statistical physics and field theory. However, for ideal,
	inviscid $\nu = 0$ flows, and in the absence of forcing, would it be
	possible to uncover a Hamiltonian description~\cite{MorrisonRMP}? 

\maketitle

The answer is delicate. Assuming incompressibility $\nabla\cdot{\bf u} = 0$ of 
the  velocity field ${\bf u}$, and unit (constant) density gives a Hamiltonian
\begin{equation}
H[{\bf u}] = \frac{1}{2}\int |{\bf u}|^2 d{\bf x} \ ,
\end{equation}  
that should lead to the three-dimensional (3D) Euler equation
through the Lie-Poisson bracket formalism where the pressure field $P$ emerges as a Lagrange multiplier enforcing 
the incompressibility constraint ~\cite{MorrisonRMP}.
In the rest of this paper, we will work in the space of periodic functions and hence solutions --- numerical and theoretical --- of 
the Euler equation will be considered on a 2$\pi$, triply-periodic cubic domain.

The Hamiltonian formulation runs into trouble when dealing with
\textit{weak} or dissipative solutions of the Euler equations: 
\begin{equation}
\frac{\partial {\bf u}}{\partial t} + {\bf u}\cdot \nabla {\bf u} = -\nabla P \ . 
\label{eq:Euler}
\end{equation}
In particular, solutions with H\"older regularity $\le \nicefrac{1}{3}$ lead to non-conservation of
kinetic energy --- the problem of anomalous dissipation in turbulence~\cite{Eyink2024} --- and
hence the breakdown of a classical Hamiltonian structure. 

Curiously, there is a form of the Euler equation which preserves its Hamiltonian structure~\cite{Lee1952} and conserves the kinetic energy. Let us chose a Galerkin wavevector ${\bf K}_{\rm G}$, such that the dynamics of the Euler 
equation is restricted between wavevectors $-{\bf K}_{\rm G} \le {\bf k} \le {\bf K}_{\rm G}$ through 
the self-adjoint, orthogonal, Hermitian Galerkin projector $\mathcal{P}_{{\bf K}_{\rm G}}$, defined as 
\begin{equation}
	{\bf v} \equiv \mathcal{P}_{{\bf K}_{\rm G}} {\bf u} = \sum_{{\bf k} = -{\bf K}_{\rm G}}^{{\bf k} = {\bf K}_{\rm G}} {\bf \hat{u}_k} e^{i {\bf k}\cdot {\bf x}},
\label{eq:PG}
\end{equation}
where ${\bf \hat{u}_k}$ are the Fourier modes of the velocity field ${\bf u}({\bf x})$ following the untruncated Euler equation~\eqref{eq:Euler}.

In what follows, it will be useful to often work with the Fourier modes ${\bf \hat{v}_k}$ of the Galerkin-truncated velocity field $\bf v(x,t)$. 
These obey, as easily derived from Eq.~\eqref{eq:Euler},  
\begin{equation}
	\frac{\partial {\bf \hat{v}_k}}{\partial t} + \sum_{\substack{{\bf p}+{\bf q}={\bf k}\\}} 
	i  \left ({\bf k}\cdot{\bf \hat{v}_p}\right ){\bf \hat{v}_q}
	= -i {\bf k} {\hat P}_{\bf k} \ ,
\label{eq:GTE_FS}
\end{equation}
with $|{\bf p}|,|{\bf q}|,|{\bf k}|\le|{\bf K}_{\rm G}|$. This, then is the Fourier-space 
representation~\cite{RayThermalReview2015} of the Galerkin-truncated, Euler equation
\begin{equation}
	\frac{\partial {\bf v}}{\partial t} + \mathcal{P}_{{\bf K}_{\rm G}}\left [{\bf v}\cdot \nabla {\bf v} + \nabla P \right ] = 0 \ ,
\label{eq:GTE_RS}
\end{equation}
with initial conditions ${\bf v}_0 =  \mathcal{P}_{{\bf K}_{\rm G}}{\bf u}_0$.
For incompressible flows, these equations are augmented by the incompressibility constraint $\bf \nabla \cdot \bf v = 0$,  which is preserved by the Galerkin projector. 

\section{Spectral signatures of thermalisation}
\begin{figure}
	\includegraphics[width = \linewidth]{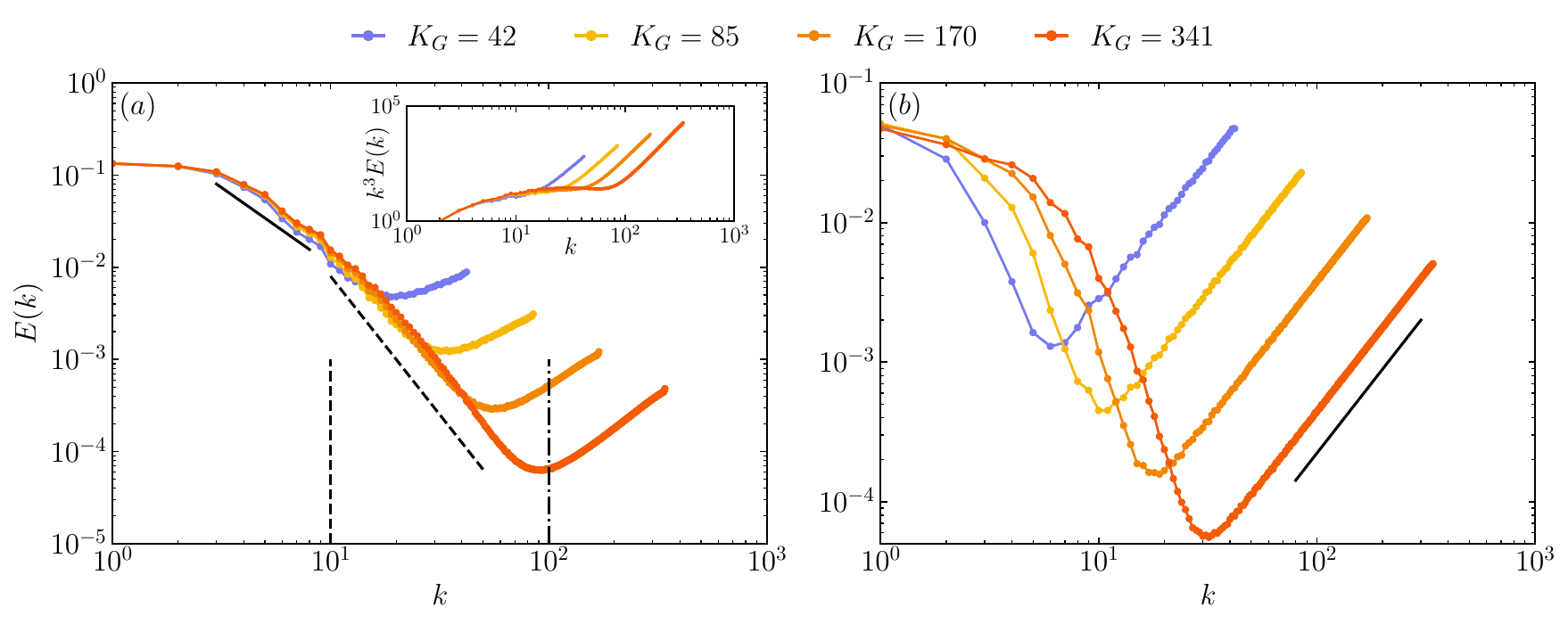}
	\caption{(a) Representative plot of the energy spectra, for different values of $K_{\rm G}$ (see legend) at $t = 3$. 
	The (left) vertical dashed line corresponds to the wavenumber $K_{\rm I}$ marking the transition from a Kolmogorov-like 
	$k^{-5/3}$ scaling  (indicated by the thick line as a guide to the eye) to one which follows a $k^{-3}$, indicated 
	by a thick and a dashed line as guides to the eye, respectively. The $k^{-3}$ scale terminates and a pre-thermalised 
	spectrum, with a positive slope, starts to develop. In the inset of panel (a), we confirm 
	the pseudo-dissipation range $k^{-3}$ scaling 
	through a compensated spectrum. At later times, such as $t = 6$ as shown in panel (b),
	the right tail follows a $k^2$ scaling as 
	indicated by the thick line. }
	\label{fig:spectra}
\end{figure}

While a superficial comparison of the Galerkin-truncated equation with the
untruncated Euler equation may suggest that the two ought to be similar, it
turns out that the solutions ${\bf u}$ and ${\bf v}$ are markedly different.
While the Euler equation, given in Eq.~\eqref{eq:Euler}, is a partial
differential equation with an infinite number of modes, its Galerkin-truncated
version is finite-dimensional, with a fixed $N_{\rm G}$ number of modes, and
thus, essentially, an ordinary differential equation.  In particular, this form
of non-viscous regularisation of the Euler equation rules out unbounded growth
of derivatives and the generation of smaller and smaller scales leading to
thermalised, chaotic  solutions~\cite{Cichowlasetal2005,Muruganetal2021} of the
Galerkin-truncated Euler equation amenable to the more conventional tools of
statistical
physics~\cite{Kraichnan1967,Kraichnan1973,Orszag1977,KraichnanChen1989}.  In
particular, the kinetic energy \rc{per unit mass} $\nicefrac{1}{2}\int |{\bf v}|^2 d{\bf x}$, and
hence the Hamiltonian, is conserved. This, along with the fact that the phase
space volume is also conserved, leads to eventual chaotic, thermalised
solutions.  Clearly the onset of thermalisation begins at the smallest scales
or the largest wavenumbers $k \gtrsim K_{\rm th}$, where $K_{\rm th}$, as
indicated by the dashed vertical line on the right in
Fig.~\ref{fig:spectra}(a), is the wavenumber (for the $K_{\rm G} = 341$
simulation spectrum), beyond which an eventual $k^2$ spectrum develops. Note,
that in panel (a), the rising spectrum on the right is still not
equipartitioned, unlike in Fig.~\ref{fig:spectra}(b), as we discuss later. With
time however, these modes $k \gtrsim K_{\rm th}$ equipartition and $K_{\rm
th}$ becomes smaller and smaller. At very long times, $K_{\rm th} \to 1$,
and the $k^2$ spectrum is the only one which remains~\cite{Cichowlasetal2005}.
At intermediate times, such as those shown in Fig.~\ref{fig:spectra}(a),
when the flow has not yet even partially thermalised (such as in Fig.~\ref{fig:spectra}(b)) but start to 
show a distinct change at large wavenumbers, the spectral behavior for modes to the left of
$K_{\rm th}$ deserves some attention.

As was shown by \citet{Cichowlasetal2005}, the large-scales of the flow, with $k \ll
K_{\rm G}$, follow a Kolmogorov scaling $k^{-5/3}$, indicated by the
dashed black lines in Fig.~\ref{fig:spectra}(a). Indeed this Kolmogorov scaling persists even when 
an equipartition spectrum kicks in as shown Fig.~\ref{fig:spectra}(b). This suggests the co-existence
of a turbulent regime and a partially thermalised small-scales. However, the
$k^{-5/3}$ spectrum gives way to a pseudo-dissipation range with $E(k) \sim
k^{-3}$ (see Fig.~\ref{fig:spectra}(a)). This is consistent with findings
reported earlier that suggest a steeper-than-Kolmogorov intermediate wavenumber
scaling. We observe that the $k^{-3}$ spectrum 
is fairly robust (see the inset of Fig.~\ref{fig:spectra}(b) with the  
compensated spectrum $k^3E(k)$) over a window of time --- the \textit{pre thermalised} phase --- 
where (i) the first signatures of truncation by way 
of a rising spectrum at large modes have appeared, and (ii) this rising spectrum is not yet equipartitioned. 
We return to this question, and its possible explanation, in a subsequent section
of this manuscript. 

This already suggests the existence of at least three time-scales. First, a cascade 
completion time $t_c$ when the nonlinear transfers have led to the largest velocity modes $K_{\rm G}$ 
being non-zero unlike its initial $t = 0$ state. This is followed by a \textit{birth} timescale, $t_b > t_c$, when the first signatures of Galerkin-truncation, namely localised damped oscillatory structures~\cite{MuruganRay2023}, 
have appeared leading to an upturned, but not equipartitioned, energy spectrum at large modes.
A third timescale, $t_{\rm th} > t_b$, corresponds to when thermalisation sets-in with a clear $k^2$ spectral 
range at the largest wavenumbers. Hence, in Fig~\ref{fig:spectra}, panel (a) corresponds to times 
$t_b \lesssim t \lesssim t_{\rm th}$ whereas panel (b) is at times $t \gtrsim t_{\rm th}$. 
In a later section, we provide a heuristic way to estimate these timescales, and in particular their 
dependence on the resolution $N$ of the simulation or, alternatively, the Galerkin wavenumber $K_{\rm G}$.

\section{Thermalisation: A brief review of the numerical studies}

Before we discuss the key results of our work, it is useful to briefly review
the numerical studies of the Galerkin-truncated Euler equation and the
understanding of thermalisation that have been developed so far.  The first
DNSs of such systems~\cite{Cichowlasetal2005}, similar to what we do here,
showed not just the phenomena of thermalisation but also that such partially
thermalised states, as shown in Fig.~\ref{fig:spectra}, could serve as a
minimal model for turbulence with a $k^{-5/3}$ scaling range with the
thermalised small-scales acting as an effective thermal bath. Indeed, studies
soon after investigated in great detail the $K_{\rm th}$ transition wavenumber
as well as developed an effective two-fluid model to explain the coexistence of
absolute equilibrium thermalised (small-scale) states with a Kolmogorov-like,
self-similar (large-scale) turbulent
phase~\cite{BosBertoglio2006,KrstulovicBrachet2008,Krstulovicetal2009,Molfettaetal2015,MKV2023}.
More recently \citet{MuruganRay2023} by using a combination of
specially curated flows --- such as vortex sheets and tubes --- along with more
generic initial conditions showed, in physical space, how thermalisation is
triggered locally in solutions of the Galerkin-truncated Euler equation
initially through small scale oscillation of wavelengths proportional to
$\nicefrac{1}{K_{\rm G}}$ which travel in directions normal to surfaces with
the largest velocity gradients. This work focussed on what we now identify as the pre thermalised phase. 
In what follows, we identify the \textit{birth}
timescale $t_b$ associated with this pre thermalised solution. Eventually, these oscillations
interact to generate other harmonics and produce patchy regions of ``white-noise'' at times corresponding to $t_{\rm th}$ which account for the $k^2$
spectral tail. 

The Galerkin-truncated 3D Euler equations are of course not the only ones which
thermalise. In fact, a more detailed understanding of how Galerkin-truncated
equations thermalise rests on a body of work which uses the one-dimensional
(1D) inviscid Burgers equation~\cite{Burgers1948} as the parent partial
differential equation. These studies were pioneered by \citet{MajdaTimofeyev2000} who showed the existence of a similar
equipartition spectrum for long time solutions of the truncated Burgers equation.
It was later discovered~\cite{RayetalTygers2011} how
shocks, integral to solutions of the inviscid partial differential equation,
lead to a resonance effect birthing spatially localised symmetric bulge-like
structures  --- christened \textit{tygers} --- at times $t_b \sim t_* - K_{\rm
G}^{-2/3}$ which eventually trigger thermalisation on timescales $t_{\rm th}
\sim K_{\rm G}^{-4/9}$~\cite{VenkataramanRay}.  ($t_*$ is the preshock time
where the inviscid Burgers equation admits it first real singularity.) Many
different aspects of these resonances, which are traced to finite-time blow-up
of the inviscid Burgers equation and the associated shocks, were studied
subsequently by several
authors~\cite{Pereiraetal2013,Leonietal2018,BrachetReview2022,Cartesetal2022,Kolluruetal2022,MKV2022}.

\section{Direct Numerical Simulations}

We perform direct numerical simulations (DNSs) of the Galerkin-truncated Euler
equation~\eqref{eq:GTE_FS} by using a fully-dealiased pseudo-spectral method
with a fourth order Runge-Kutta scheme for time-marching on triply-periodic
cube of length  $2\pi$. We consider five different resolutions with
collocation points $N^3 = 64^3, 128^3, 256^3, 512^3$ and $1024^3$. This allows us to
explore 5 different values of the Galerkin-truncation wavenumber, $21 \le
K_{\rm G} \le 341$. The time-step $10^{-3} \le \delta t \le 10^{-4}$
is chosen separately for each resolution to ensure numerical stability. We choose an
initial (incompressible) velocity field ${\bf v}_0$ such that kinetic energy is concentrated at the
largest scales at $t = 0$. Our numerical scheme and choice of $\delta t$ ensures that the
initial kinetic energy is conserved for all times. 

Somewhat paradoxically --- given the relative importance of the Euler equation ---  and unlike the 1D Burgers equation, 
several questions remain open. Some of them, which  relate to the onset of thermalisation and indeed the
pre and partially thermalised phase illustrated in Figs.~\ref{fig:spectra}(a) and (b), respectively,  are now addressed over 
the next two sections. 

\section{The $k^{-3}$ spectrum and a constant Flux}

We first investigate whether the self-similar form of the spectrum for
$k \lesssim K_{\rm th}$ is truly an \textit{``inertial''} range associated
with a constant energy flux $\Pi$. We also seek possible explanations of the origin of
the pseudo-dissipation range $k^{-3}$ scaling, for $K_{\rm I} \le k \le K_{\rm G}$, as seen in Fig.~\ref{fig:spectra}(a).
It is important to caveat this by stressing that there is no \textit{true}
dissipation in the Galerkin-truncated Euler equation. It is a conservative
system with a Hamiltonian description and conserves energy exactly. The
thermalised scales do not dissipate but merely act as a thermal bath for the
larger scales to cascade its energy into. Our use of the word pseudo-dissipation
range is sociological and borrows from the terminology used in fully developed
turbulence for the transition scales between a $k^{-5/3}$ (up to intermittency
corrections) inertial range and the true exponential dissipation range.

Could the $k^{-3}$ scaling be universal for the pre-thermalised phase, and where
could it stem from? A possible, admittedly heuristic, explanation for this may
lie in the eddy-viscosity ideas of the two-fluid model proposed by \citet{Krstulovicetal2009}. Let $E(k)$ be the yet undetermined
energy spectrum lying between the $k^{-5/3}$ (turbulence-like) and $k^2$
(thermalised) scales at wavenumbers $K_{\rm I} \le k \le K_{\rm G}$.
Dimensionally, and ignoring constants, the \textit{typical} velocities relate
to this spectrum as ${\hat v}^2_k \sim kE(k)$. An  eddy-viscosity argument
leads to an eddy-damping timescale $1/\tau_d \sim \nu_{\rm eff} k^2$, where
$\nu_{\rm eff}$ is a phenomenological coefficient of eddy viscosity. This leads
to an effective dissipation $\epsilon \sim \nu_{\rm eff} k^2 {\hat v}_k^2$
which ought to be equal to a constant, so far hypothetical,  energy flux $\Pi$
that cascades from the large scale kinetic energy to the small-scale, thermalised
bath. Hence, $\Pi = \nu_{\rm eff} k^3 E(k)$, giving way to the $k^{-3}$ scaling of
the pseudo-dissipation range.

\begin{figure}
	\includegraphics[width = \linewidth]{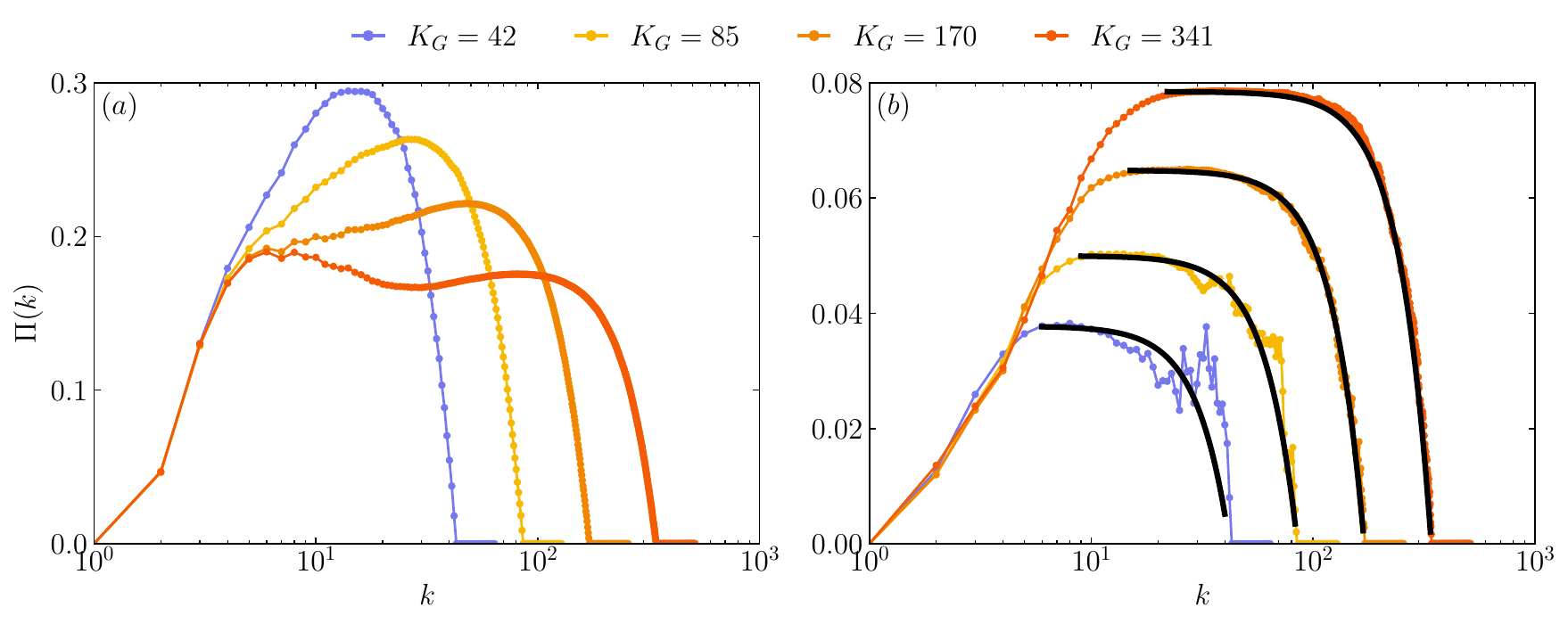}
	\caption{Representative plots of the energy flux, for different values of $K_{\rm G}$ (see legend), at (a) $t = 3$ and 
	(b) $t = 6$. The solid black curve,  which overlay the data points in panel (b), 
	are obtained from a theoretical estimate of $\Pi$ (see text) and provide a reasonable approximate of the flux for large $k$ and $K_{\rm G}$.}
	\label{fig:flux}
\end{figure}

Such an argument, while compelling, has several drawbacks. Most importantly, the idea 
of an eddy viscosity for the pre-thermalised phase and not for the partially thermalised phase,
where we know the pseudo-dissipation range scaling is steeper than -3  (see Fig~\ref{fig:spectra}(b)),  
is somewhat puzzling in the absence of any strong argument to show relatively dominant Reynolds stresses 
at timescales $t_b \lesssim t \lesssim t_{\rm th}$. Furthermore, the assumption of constant fluxes 
seems marginally true. While for our largest simulations this constant flux is perhaps a persuasive 
argument (see Fig.~\ref{fig:flux}(a)), the evidence for this is somewhat lacking for the lower resolutions 
which nevertheless display the $k^{-3}$ scaling (see inset in Fig.~\ref{fig:spectra}(a)). Hence we caution the 
reader about the pitfalls of the argument we use to derive the numerically robust, pre-thermalised phase, 
pseudo-dissipation range scaling, and a fuller theoretical understanding of this is warranted in future.

One of the assumptions made above to understand the spectral scaling is that of a constant flux $\Pi$ across the
$k^{-5/3}$ and $k^{-3}$ scaling ranges. Is this really consistent with
measurements from our DNSs? In Fig.~\ref{fig:flux} we show $\Pi$ for all our
resolutions and at the same times for which we had shown the energy spectra in
Fig.~\ref{fig:spectra}. At late times $t > t_{\rm th}$ we observe (Fig.~\ref{fig:flux}(b)) 
a remarkably robust, flat, $k$-independent plateau, indicating effectively an \textit{inertial} range 
at scales larger than the thermalised scales. However at times $t_b \lesssim t \lesssim t_{\rm th}$, 
as seen in (Fig.~\ref{fig:flux}(a)) and discussed in the previous paragraph, this constant flux regime 
is less compelling. 

Energy conservation allows us, nevertheless, to estimate the theoretical form
of the flux at least in the ultraviolet regime. Consider the partially
thermalised spectral form to be $E(k) = C k^2$, where $C$ is a time-dependent
constant whose value can be estimated numerically. The nonlinear energy
transfer $T (k) = \frac{dC}{dt}k^2$ relates trivially to the time variation of
this spectrum, and which, in turn relates to the energy flux $\Pi (k) =
\sum_{k^\prime = k}^{K_{\rm G}}  T(k')\approx \nicefrac{1}{3}\frac{dC}{dt}\left [
	K_{\rm G}^3 - k^3\right ]$. In Fig.~\ref{fig:flux}(b), we overlay on
our DNSs data this theoretical curve (with $\frac{dC}{dt}$ obtained from the
spectra measurements) and find a convincing match with the flux
measurements from our simulations. However, at earlier times, such as the one shown in
Fig.~\ref{fig:flux}(a), this theoretical curve cannot predict the lack of a clear plateau at intermediate $k$. 

\section{The Time-scales for Thermalisation}

We now address the question of the various timescales in the dynamics of
thermalisation and how they depend on $K_{\rm G}$. We recall that this question
was answered for the 1D Burgers equation, and the timescale $t_b$ and $t_{\rm
th}$ were estimated semi-analytically through the idea of
resonances~\cite{RayetalTygers2011} for the former, and Reynolds
stresses~\cite{VenkataramanRay} for the latter. Unfortunately, such tools are
not available at our disposal for the question of thermalisation in the Euler
equation~\cite{MuruganRay2023}. Furthermore, in this problem we also deal with
the cascade-completion time $t_c$, the analog of which for the Navier-Stokes
equation is a well studied
problem~\cite{Brachetetal1983,HerringKerr1993,RayetalEPJB}.

Let us begin with the cascade completion time $t_c$.  Given the initial
localisation of the kinetic energy in the first few wavenumbers $1 \leq k
\lesssim k_0$, where $k_0 \ll K_{\rm G}$, the nonlinear advection leads to
transfer of energy (and momentum) to wavenumbers $k > k_0$. We should thus be
able to construct a \textit{cascade} time scale $t_c$ which is a measure of the
time taken for the largest wavenumbers around $K_{\rm G}$ to be excited and
have finite, non-zero energies.  

One way to estimate $t_c$ is through the time evolution of the enstrophy
$\Omega = \sum k^2 E(k)$. Given the sensitivity of the enstrophy to small-scale
excitations, $\Omega$ grows in time as the energy cascade kicks in and higher
wavenumbers get excited. This process is consistent with what is known for
(decaying) simulations of the Navier-Stokes equation as well~\cite{Brachetetal1983}. However, there is
a critical difference. In the Navier-Stokes, $\Omega$ reaches a maximum at
$t_c$ and subsequently decays with time as the total energy of the system
decays in the absence of
forcing~\cite{Brachetetal1983,HerringKerr1993,RayetalEPJB}. For our problem
this should not happen because of energy conservation and we ought to expect a
growth and saturation of $\Omega \to \Omega_\infty$ at late times.

\begin{figure}
	\includegraphics[width = \linewidth]{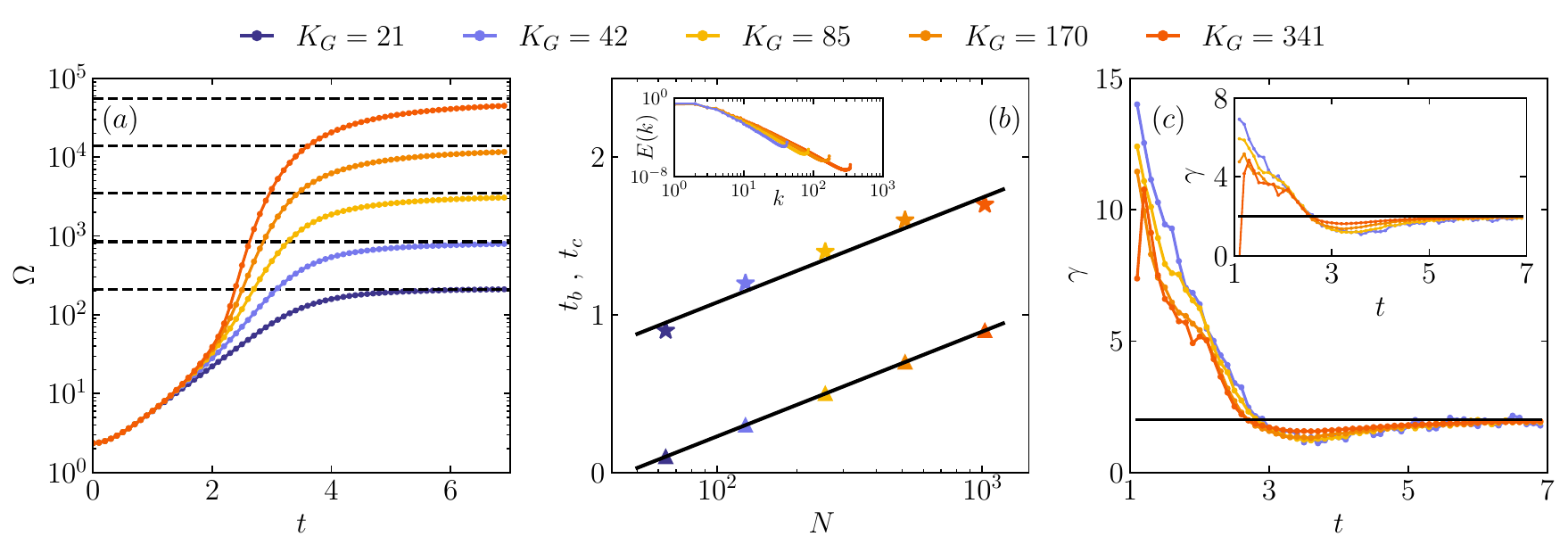}

	\caption{(a) A plot of the enstrophy $\Omega$ versus time for different Galerkin-truncation wavenumbers $K_{\rm G}$ showing the growth 
	and $K_{\rm G}$ dependent saturation at $\Omega_\infty$, indicated by the horizontal dashed lines (see text). At short times the curves 
	for different $K_{\rm G}$ collapse, and as smaller scales get excited, they separate.
	(b) The timescales for cascade completion $t_c$ (triangles), and the \textit{birth} $t_b$ (stars) of the localised oscillatory structures which eventually trigger thermalisation plotted against the logarithm of $N$. The linear curves suggest that both these timescales depend 
	logarithmically on $N$. The inset shows the energy spectrum, for different $N$, or equivalently for different $K_{\rm G}$, at $t_b$ showing the departure at the ultraviolet end from 
	the $k^{-3}$ scaling before the onset of an equipartition $k^2$ spectrum. (c) The variation of the spectral slope for wavenumbers in the 
	\textit{boundary} layer $K_{\rm BL} = 10\%$. The horizontal dashed line indicates a slope of 2 corresponding to the 
	equipartition $E(k) \sim k^2$ spectrum. We note a $K_{\rm G}$-independent convergence to the thermalised regime suggesting that the 
	thermalisation time $t_{\rm th}$ is, unlike the 1D Burgers problem as well as $t_c$ and $t_b$, independent of $K_{\rm G}$. In the inset we 
	show the analogous result for $K_{\rm BL} = 15\%$ leading to the same conclusion.}
	\label{fig:timescale}
\end{figure}

In Fig.~\ref{fig:timescale}(a) we show the time evolution of $\Omega$ for 
different resolutions.  Several things stand out. At short times the
curves collapse on top of each other till they start separating from the lowest
resolution onward. At long times the curves saturate to values
$\Omega_{\infty}$ which depend on $K_{\rm G}$. 

It is possible to theoretically estimate $\Omega_{\infty}$. When the 
flow is fully thermalised, the energy spectrum follows $E(k) = C_\infty k^2$, which, given the 
conserved initial kinetic energy $E_0$, allows for the determination 
$C_\infty = 6E_0 [(K_{\rm G})(K_{\rm G}+1)(2K_{\rm G} + 1)]^{-1} \approx 3E_0/K_{\rm G}^3$, for large $K_{\rm G}$. 
By using this, it is trivial to show that, for large $K_{\rm G}$,  the 
saturation value $\Omega_{\infty} =  \nicefrac{3}{5}E_0 K_{\rm G}^2$. In Fig.~\ref{fig:timescale}(a) we plot 
$\Omega_{\infty}$, as dashed horizontal lines, to see how the system evolves to 
the fully thermalisation phase.  

Unlike the Navier-Stokes equation, $\Omega$ is not an ideal 
metric for estimating $t_c$. Of course, one possibility is to estimate the time when 
the curves for different values of $N$ separate from the collapsed curves at short times as 
seen in Fig.~\ref{fig:timescale}(a). Consistent with intuition, a time-scale obtained from this observation 
grows with $N$. However, there is a drawback to using this prescription: It does not allow us to determine 
$t_c$ for the largest resolution since this measurement is, by definition, relative to the $\Omega$ curve 
for $N = 1024$. 

A more natural way to think about $t_c$ is to estimate the time when the kinetic
energy for the largest mode is non-zero. This would correspond to a 
culmination of the cascade process whence all scales have been excited.
Numerically, for $t < t_c$, the energy of the largest mode in the kinetic
energy spectrum $E(K_{\rm G})$ is below the machine precision value. We now
record the time $t_c$ when this threshold is crossed. Of course the estimation of $t_c$
is somewhat inaccurate in simulations where the data is saved at discrete time intervals. Nevertheless, within such an approximate approach, we evaluate the
cascade-completion time-scale $t_c$, and find, in Fig.~\ref{fig:timescale}(b), a
logarithmic scaling of $t_c \sim \log N$. 

The cascade completion time $t_c$ does not indicate the onset time for thermalisation, $t_{\rm th}$, or indeed the time $t_b$ at which the precursor to thermalisations show up:
$t_c < t_b < t_{\rm th}$. \citet{MuruganRay2023} had shown that
for the truncated Euler equations the first signs of the solution being
\textit{different} from $\nu \to 0$ Navier-Stokes solutions are localised,
monochromatic oscillatory structures which emerge \textit{out of the blue} in
physical locations proximate to structures with the steepest gradients. These
structures, constituting the pre-thermalised phase, are \textit{born} at times $t_b$ and eventually trigger
thermalisation with the energy for modes beyond $K_{\rm th}$
equipartioned at times $t \gtrsim t_{\rm th}$. This two-step trigger, resulting in a dual timescale,
is reminiscent of thermalisation
in the one-dimensional Galerkin-truncated Burgers equation. 

We test this idea on the data from our DNSs. Careful observations suggest that
at times soon after $t_c$, well before the first signs of an incipient
equipartition spectrum is detectable, the $k^{-3}$ scaling extends all the way
to $K_{\rm G}$. In time, the ultra-violet end of the spectrum starts to bend
upward, but not yet equipartitioned, over a time window $t_b \leq t \lesssim
t_{\rm th}$ constituting the pre thermalised phase.  We see an example of this
in the inset of Fig.~\ref{fig:timescale}(b). 

A simplistic formulation of when the $k^{-3}$ gives way to a rising
spectrum in the pre thermalised phase is to estimate
the time $t_b$ at which the energy of the largest mode $E(K_{\rm G}) \gtrsim C
K_{\rm G}^{-3}$. The time-dependent constant $C$ is
obtained from our data.  We find from such measurements that $t_b \sim
\log N$, illustrated in Fig.~\ref{fig:timescale}(b), just like the scaling form for $t_c$, but with a 
different prefactor since $t_b > t_c$.

We finally consider the thermalisation timescale $t_{\rm th}$. Unlike the
resonance conditions and consequent Reynolds stress triggered collapse from a
monochromatic to white noise form for the 1D Burgers equation, theoretical
progress is difficult in \rc{the Galerkin-truncated Euler} problem. We therefore resort to the following
ad-hoc measure.  

We consider a boundary layer of wavenumbers, lying between $K_{\rm BL} \le k \le K_{\rm G}$, 
such that its extent $K_{\rm G} - K_{\rm BL}$ is a percentage $\mathcal{X}$ of the Galerkin-truncation 
wavenumber $K_{\rm G}$. If we measure the spectral slope of $E(k) \sim k^\gamma$ within this boundary layer, $\gamma \to 2$ as 
$t \to t_{\rm th}$. Of course, it is neither obvious nor necessary (for this calculation) that the boundary layer spectrum is 
necessarily a power-law; what matters is that as the boundary layer thermalises it ought to scale as $E(k) \sim k^2$.  

In Fig.~\ref{fig:timescale}(c), we show a plot of $\gamma$ vs time, with
$\mathcal{X} = 10\%$, for different values of $K_{\rm G}$. The $\gamma \neq 2$
range is irrelevant for the reason discussed above. What is important is how
$\gamma$ approaches 2 for different $K_{\rm G}$. Within statistical errors in the
measurement of $\gamma$, our results suggest that $t_{\rm th}$ is independent
of or only weakly dependent on $K_{\rm G}$. This is unlike what is known for
the analogous  problem in the 1D Burgers equation \cite{VenkataramanRay}. 

It useful to add a caveat here. While ideally, the time for the onset of
thermalisation ought to be estimated with as small a percentage $\mathcal{X}$
as possible, a numerically trustable fit over a self-similar spectrum
constrains us on the smallest  $\mathcal{X}$ that we can practically use. 
Thus we confirm the conclusion $t_{th} \sim K_{\rm G}^0$ by varying
$5\% \le \mathcal{X} \le 20\%$; in the inset of Fig.~\ref{fig:timescale}(c) we
show a plot of $\gamma$ for $\mathcal{X} =  15\%$ which leads to same conclusion as before.

\section{Outlook \& Perspective}

As we conclude, let us address the question of the relevance of such studies 
for turbulence and statistical physics. While a detailed review of this
is beyond the scope of the present paper, let us offer some perspective
on this question.  

The nature of thermalised solutions would lead one to believe that these
systems have very little to do with turbulence. This is only partially true.
For problems more directly relevant to natural or experimental turbulent flows,
one of the successes of such solutions has been in providing 
\textit{an} explanation~\cite{FrischetalBottleneck2008,FrischetalBottleneck2013} 
for the bottleneck phenomenon~\cite{Falkovich1994,Lohse1995}. A bump between the
inertial and dissipative ranges in the compensated $k^{5/3} E(k)$ energy
spectrum in turbulence~\cite{Pak1991,SheJackson1993,Gotohetal2002,Kurien2004,Verma2007,Mininni2008,Ishihara2009,Donzis2010}. 
Numerical simulations have confirmed how such effects
are exaggerated with the use of hyperviscosity~\cite{FrischetalBottleneck2008,BanerjeeRayHyperviscosity2014} 
instead of the regular Laplacian
dissipative term.  Such explanations lie in the observation that 
higher-order hyperviscous terms will converge to the Galerkin-truncated 
equation~\cite{FrischetalBottleneck2008,BanerjeeRayHyperviscosity2014} and hence lead to the non-monotonic
energy pile up --- bottleneck --- between the inertial and far dissipation ranges.
Likewise, a more generalised form of the Galerkin projection --- the so-called
decimation projector~\cite{FrischetalDecimation2012} --- have been used to
address questions of
intermittency~\cite{Lanotte2015,Buzzicottietal2016Burgers,Buzzicotti2016NS} and
irreversibility~\cite{Ray2018,Picardoetal2020} which remain central to the
statistical physics and statistical field theory of turbulence. These ideas have been further 
supplemented by studies related to entropy in the partial and fully thermalised 
regimes of such equations~\cite{MKV2024}.

A more involved aspect of studies of Galerkin-truncated systems comes from numerical
investigations for potential finite-time singularities in the Euler equation
which inevitably tackle such questions. This is because a spectral
(numerical) approach to solving equations such as the Navier-Stokes for
turbulence naturally lead to solving the Galerkin-truncated equation when
$\nu = 0$. Hence, as reported by \citet{BustamanteBrachet2012}, and more recently for the axis-symmetric problem by Pandit and collaborators~\cite{Kolluruetal2022,KolluruPandit2024,KolluruetalJCP2024},
singularity tracking through techniques such as the analyticity strip
method~\cite{Sulemetal1983} would invariably encounter the effects of
thermalisation. Indeed, this provides a key motivation and so far an open
problem of how to circumvent or indeed delay the onset of thermalisation and 
recover possible dissipative solutions of the Euler equation.  So
far, such attempts have been restricted to very special cases for the Euler
equation~\cite{MuruganRay2023,Fehnetal2022}. This is unlike the 1D Burgers
equation where a more concrete theory of how such questions can be tackled have
been reported~\cite{Muruganetal2020,Kolluruetal2024}. We hope that our work on
the dynamics of thermalisation may spur more efforts in such directions. One possibility 
would be to identify in a coarse-grained manner, such as the one developed recently for the Navier-Stokes 
multifractality problem~\cite{Mukherjeeetal2024}, how inhomogeneous is the distribution of the trigger points of thermalisation. This 
may well lead to a numerical prescription on how these spots can be selectively 
eliminated.
Finally, for the Galerkin-truncated Euler equation to serve as a sub-grid scale model 
for turbulence, inertial range intermittency through measurements of the equal-time scaling exponents, 
for example, in the $k^{-5/3}$ scaling range in the partially thermalised phase, needs further 
investigation. 

\begin{acknowledgments}
The authors declare that they have no competing interests.
SSR acknowledges the Indo–French Centre for the Promotion of Advanced
Scientific Research (IFCPAR/CEFIPRA, project no. 6704-1) for support. The simulations were performed on the ICTS Contra cluster. R,
MSK, and and SSR acknowledge the support of the DAE, Government of India, under projects nos. 12-R\&D-TFR-5.10-1100 and RTI4001.
The authors thank R. Mukherjee and S. Sahoo for several useful discussions. R thanks A. Sherry, J. Kethepalli, and T. Ray for their insights. The code used to generate the data is hosted publicly on \href{https://github.com/Rajarshi-prime/GT-Euler}{GitHub}. Microsoft and GitHub copilot were used for code development and plot formatting. 
The data from our numerical simulations are available on request.
\end{acknowledgments}
\bibliography{references}  
\end{document}